\documentclass[12pt,epsfig,fleqn]{article}
\usepackage{amsmath,amssymb,amsfonts,amsbsy}
\usepackage{graphicx}
\usepackage{wrapfig}

 \newcommand{\be}{\begin{eqnarray}}
 \newcommand{\ee}{\end{eqnarray}}
 \newcommand{\nee}{\nonumber\end{eqnarray}}
 \newcommand{\nn}{\nonumber\\}

  \newcommand{\bc}{\begin{center}}
 \newcommand{\ec}{\end{center}}
  \def\a               {\alpha}

\def\m             {\mu}

\def\s              {\sigma}
\def\g              {\gamma}
\def\d             {\delta}
 
\begin{document}

 \bc
 { \bf The difference cross sections of unpolarized SIDIS  with transverse momentum dependence}
\ec
\vspace{5mm}

\bc
{\bf Ekaterina Christova}
\vspace{.1cm}

{\it Institute for Nuclear Research and Nuclear Energy, Sofia, Bulgaria}
\ec
\vspace{.1cm}

\begin{abstract}

 Previously we showed that, based only on C and SU(2) invariance,  the difference
cross sections of  hadrons with opposite charge in SIDIS $e+N\to l+h+X$ is expressed solely
in terms of the valence-quark densities and certain non-singlet combinations of FFs.
This allowed to determine these quantities in a model independent way. Now we extend this approach
to processes when the transverse momentum of the final hadron is measured as well.
We show that the difference cross sections of unpolarized SIDIS on proton and deuterium targets, 
 $d\s_N^{h^+-h^-},d\s_N^{\pi^+-\pi^-}$ and $d\s_N^{K^+-K^-}$,
are expressed solely in terms of the TMD unpolarized valence quark densities and FFs,
and the valence-quark Boer-Mulders and Collins functions. This  allows
to determine them separately and study flavour dependence of the quark  transverse  momentum. 
Measurements on deuterium target, $d\s_d^{h^+-h^-},d\s_d^{\pi^+-\pi^-}$ and $d\s_d^{K^+-K^-}$,
 provide 3 independent measurements for the sum of the TMD valence quark densities and Boer-Mulders functions:
$(u_{1,V}+d_{1,V})$ and $(h_{1,uV}^\perp + h_{1,dV}^\perp )$.
\end{abstract}

\section{Introduction}

Now it seems quite well established that the simple collinear picture of the quark-parton model appears
 too simple to explain existing experimental data. The measured azimuthal asymmetries in the direction of the final hadrons
 shows that the transverse momentum of the quarks should be necessarily taken into account.
 This leads to the following 3 main differences as compared to the collinear parton model:
  1) the known parton densities (PDFs) and fragmentation functions (FFs)
 depend not only on the longitudinal, but on the transverse momenta of the quarks as well -- we start to deal with
  transverse momentum dependent parton densities (TMD-PDFs)
 and fragmentation functions (TMD-FFs), 2) new type of TMD parton densities  and FFs arise
 from correlations among the transverse components of quark momentum or spin,
 and the longitudinal components of the particles in the process and, 3) the TMD-PDFs
 and TMD-FFs always enter the cross sections in convolutions over the quark transverse momenta.

 This makes the problem of extracting the transverse momentum densities and FFs from experiment
  considerably more complicated.
 In order to simplify analysis lot of assumptions on the TMD-functions,
in addition to those on the collinear  PDFs and FFs, are made for the transverse
  momentum dependence: it is factorized,
it is flavour blind, it is hadron blind, etc.  Though sometimes quite reasonable, these
are  ad hoc model assumptions, motivated mainly by simplicity and
 do not follow from QCD theory of strong interactions and thus, introduce uncontrolled uncertainties.

 For these reasons it is important to find  measurable quantities, that would
   extract  TMD functions  without or with less additional
  assumptions.

    Previously this task was fulfilled
  for the  collinear polarized PDFs. We showed \cite{we_assym} that,
   based only on charge conjugation (C) and iso-spin SU(2)-invariance of strong interactions,  the so called "difference
  asymmetries" in semi inclusive deep inelastic scattering (SIDIS) of longitudinally polarized leptons on longitudinally polarized nucleons
  determine the polarized valence-quark PDFs  in a model independent way. Such measurements were
  fulfilled and the polarized valence-quark densities were determined directly \cite{CERN}.

   Later, the same approach was used for the collinear FFs.  We showed \cite{we_FF} that
differences between the cross-sections  for producing  hadrons and their
  antiparticles in unpolarized SIDIS,
  allow to determine non-singlet combinations of the collinear FFs in a model independent way and test most of the
  commonly used assumptions. Recently, this approach was applied to HERMESS data and
  the non-singlet combination of the pion fragmentation functions was determined with very good precision \cite{LSS}.

  Now we extend this approach to the non-collinear picture of the parton model, when
  parton densities and fragmentation functions depend on the transverse momentum of the quarks as well.
     Transverse momentum of the quarks  plays
    crucial role when not only the energy, but the transverse momentum of the final hadron is measured.

 In this paper we consider unpolarized  SIDIS and show how, based only
  on the general symmetries of C and SU(2) invariance,
   information on certain combinations of the TMD-PDFs and TMD-FFs can be obtained in a model independent
 way. The key experimental
  ingredients are the differences between cross-sections for producing  hadrons and producing their
  antiparticles, i.e. data on $ d\sigma^{h-\bar{h}} \equiv d\sigma^h - d\sigma^{\bar{h}} $, for $h=h^\pm, \, \pi^\pm,\,K^\pm$.

The paper is organized as follows. In the  next section we recall the general expression
 of the cross section and introduce the notation. In Sections 3, 4 and 5 we give the difference cross sections
 for any charged hadrons, for charged pions and charged kaons, respectively, Sect. 5 ends up
 with a brief summary of the obtained results
 for charged hadron production.  In Sect. 6 we present
 the difference cross section for charged and neutral kaons. In all cases
 we present the results for proton and deuterium targets.
  In Sect.7 we discuss the standard parametrizations and those appropriate to the
  considered approach and, the possibilities to study flavour and $Q^2$ dependence in
  the quark transverse momenta in the TMD's. We end up
 with our comments and conclusions.

\section{The  cross section -- general expression}

 The cross section for SIDIS of unpolarized leptons $l$ on unpolarized nucleons $N$:
 \be
 l\,(l^\m )+ N\,(P^\m ) \to l'\,({l'}^\m ) + h\,(P_h^\m ) +X
 \ee
   exhibits a characteristic $\cos 2\phi_h$ and $\cos\phi_h$
 azimuthal dependence in the  kinematic region of low $P_T\simeq
\Lambda_{QCD}\ll Q$,  $\phi_h$ is the azimuthal angle of the produced hadron
 $h$.
The general expression for the cross section in the TMD factorization scheme \cite{fact},
 in the one-photon exchange approximation and in LO of QCD,
 reads \cite{Bacchetta,general}:
 \be
 \frac{d^5\s_p^h}{dx_B\,dQ^2\,dz_h\,d^2{\bf P}_T}&=&\frac{2\pi
\a^2_{em}}{Q^4}\left\{[1+(1-y)^2]\,F_{UU}^h+ 2\,(1-y)\,\cos
2\phi_h\,F_{UU}^{\cos 2\phi ,h}+\right.\nn
&&\left.+2\,(2-y)\sqrt{1-y}\,\cos \phi_h\,F_{UU}^{\cos
\phi_h,h}\right\}.\label{1}
\ee
 Here   ${\bf P}_T$ is the transverse momentum of the final hadron  in the $\g^*-p$ c.m. frame,
 $x_B,\,z_h,\,Q^2$ and $y$ are the usual measurable SIDIS quantities:
\be
x_B=\frac{Q^2}{2(P.q)},\quad z_h=\frac {(P.P_h)}{(P.q)},\quad Q^2=-q^2,\quad y=\frac{(P.q)}{(P.l)},\quad q=l-l'.
\ee

 Throughout the paper we use the kinematic configuration and the results of \cite{general}.
  However, we write the $F_{UU}$'s in a slightly different form --
we indicate explicitly the indicies   of the quark flavours, $q=u,\bar u , d,\bar d,s,\bar s$ and the type of the
produced hadron $h$, and we single out the quantities that are flavour and hadron-type independent.
  We have:
\be
F_{UU}^h&=&\sum_q e_q^2\,f_{1q}\otimes D_{1q}^h\nn
F_{UU}^{\cos 2\phi ,h}&=&\sum_q e_q^2\left[h_{1q}^\perp \otimes H_{1q}^{\perp ,h} \otimes w_2^\perp +\frac{2}{Q^2}
\,f_{1q} \otimes D_{1q}^h \otimes w_2\right]\nn
F_{UU}^{\cos \phi ,h}&=&-\frac{2}{Q}\,\sum_q e_q^2\left[h_{1q}^\perp \otimes H_{1q}^{\perp ,h}\otimes w_1^\perp +
\,f_{1q} \otimes D_{1q}^h\otimes w_1\right],\label{FUU}
\ee
where the convolutions are defined as:
\be
f\otimes D \otimes w=\int d^2{\bf k}_\perp d^2{\bf p}_\perp
\d^2({\bf P}_T-z_h{\bf k}_\perp -{\bf p}_\perp )f(x_B,k_\perp )D(z_h,p_\perp )w({\bf P}_T ,{\bf k}_\perp ).\nn
\ee
Here  $\bf k_\perp$
is the transverse momentum of the quark in the target nucleon, $k_\perp = \vert \bf k_\perp\vert$;
$\bf p_\perp$ is the transverse momentum of the final hadron with respect to the direction of the fragmenting quark,
$p_\perp = \vert \bf p_\perp\vert$;
  at the order
   $(k_\perp /Q)$, for the measured transverse momentum of the final hadron, we have ${\bf P}_T=z_h {\bf k_\perp}+{\bf p_\perp}$.

The  functions $w_i$ and $w_i^\perp$ are flavour and hadron-type independent, and contain only kinematic factors:
\be
w_1&=&(\hat {\bf P}_T{\bf k}_\perp )\nn
w_2&=&2(\hat {\bf P}_T{\bf k}_\perp )^2-k_\perp^2\nn
w_1^\perp &=&\frac{k_\perp^2\left(P_T-z_h(\hat {\bf P}_T{\bf k}_\perp )\right)}{z_hM_hM}\nn
w_2^\perp&=&\frac{({\bf P}_T{\bf k}_\perp )-2z_h(\hat {\bf P}_T{\bf k}_\perp )^2+z_h\,k_\perp^2}{z_hM_hM}\nn
\hat {\bf P}_T&=& \frac{{\bf P}_T}{\vert {\bf P}_T\vert},\qquad P_T=\vert {\bf P}_T\vert
\ee
The only dependence of $w_i$ and $w_i^\perp$ on the final hadron
$h$ is through $M_h$. However this  is be irrelevant for us, as we shall consider
the production of $h$ and its anti-particle $\bar h$, for which $M_h=M_{\bar h}$.

 In (\ref{FUU}) $f_{1q}(x,k_\perp )$ and $D_{1q}^h(z, p_\perp )$ are the unpolarized TMD
  parton distribution and fragmentation functions respectively,
  $h_{1q}^\perp (x, k_\perp )$ are the  Boer-Mulders distribution functions \cite{BM} that describe the probability to find a
transversely
polarized quark $q$ in an unpolarized proton, $H_{1,q}^{\perp, h}(z, p_\perp )$ are the Collins
fragmentation functions \cite{Collins},
 that describe the probability for a transversely polarized quark $q$ to produce an unpolarized hadron $h$
 with  a fraction $z$ of the longitudinal momentum  and transverse momentum $p_\perp$ with respect
   to the momentum of the fragmenting quark.

 The first term in (\ref{1}), with $F_{UU}^h$, describes the $\phi_h$ independent cross section, it is expressed through
$f_{1q}(x, k_\perp )$ and $D_{1q}^h (z, p_\perp )$.

Two mechanisms generate the azimuthal $\cos\phi_h$ and $\cos 2\phi_h$-dependence:

-- the Cahn effect~\cite{Cahn}, which is a purely kinematic effect, generated by the intrinsic transverse quark momenta.
 It is described by
the unpolarized TMD functions $f_{1q}$ and $D_{1,q}^h$, and is a sub-leading effect: $1/Q^2$ contribution to
 $F_{UU}^{\cos 2\phi_h}$ and $1/Q$ contribution to
 $F_{UU}^{\cos \phi_h}$.

-- the Boer-Mulders effect~\cite{BM}, which points to the existence of non-zero transverse polarization of the quarks
  and is described by the TMD functions with transversely polarized quarks: $h_{1q}^\perp $
 and $H_{1q}^\perp $. The induced $\cos 2\phi_h$-dependence is a leading (twist-2) effect -- the first term in
$F_{UU}^{\cos 2\phi_h}$ in eq. (\ref{FUU}), the  $\cos \phi_h$-dependence is a sub-leading ~ $1/Q$ effect.

 Note that in (\ref{FUU}) we have included the Cahn
 contribution  to the $\cos 2\phi_h$-term, though it is of higher $1/Q^2$-order and the other
 terms of the same order are not included.  We think this gives a more clear physical
 picture of the different contributions, but it is irrelevant for the discussions in the paper.


 \section{The difference cross section with  ${\bf h^\pm}$}

Here we shall consider the difference of the cross sections for producing a hadron $h$ and its anti-particle $\bar h$,
 when the type of the hadrons is not specified and they distinguished only by their charge:
\be
d\s_N^{h-\bar h}\equiv \frac{d^5\s_N^h}{dx_B\,dQ^2\,dz_h\,d^2{\bf P}_T}-
\frac{d^5\s_N^{\bar h}}{dx_B\,dQ^2\,dz_h\,d^2{\bf P}_T}\label{diff}
\ee
where $N$ stands for a proton or a neutron target, $N=p,n$.

Charge conjugation invariance of strong interactions implies the following relations on the
unpolarized TMD and Collins FFs:
 \be
D_{1,q}^h=D_{1,\bar q}^{\bar h}, \qquad  D_{1,\bar q}^h=D_{1, q}^{\bar h},\quad
H_{1,q}^{\perp ,h}=H_{1,\bar q}^{\perp ,\bar h},\quad
H_{1,\bar q}^{\perp ,h}=H_{1, q}^{\perp ,\bar h}\label{C}
\ee
Using these relations, from (\ref{1}) and (\ref{diff}), we obtain  the difference cross section $d\s_N^{h-\bar h}$.
 It is easily shown that the
  azimuthal dependence in $d\s_N^{h-\bar h}$ remains the same as in $d\s^h_N$,
  but the expressions for $F_{UU}^{h-\bar h}$ considerably simplify. We show that,
  based only on the  general  properties of charge conjugation invariance of strong interactions,
only the contributions of the largest TMD valence quark densities
  survive in $F_{UU}^{h-\bar h}$, $F_{UU}^{\cos 2\phi ,h-\bar h}$
 and $F_{UU}^{\cos \phi ,h-\bar h}$. Bellow we give  the expressions for
 the $F_{UU}^{h-\bar h}$'s for proton and deuterium targets separately.

  As usual, subindex 1 indicates
for a transverse momentum dependence:
$f_{1q}(x,k_\perp) \equiv q_1 (x,k_\perp)$ and so on.

 \subsection{ on proton target}

 The expression for the difference cross section on a proton target $d\s_p^{h-\bar h}$ is analogous to $d\s_p^{h}$, (\ref{1}),
  in which $F_{UU}^h$ are replaced by the corresponding $F_{UU}^{h-\bar h}$ as given bellow:
\be
F_{UU}^{h-\bar h}&=& e_u^2\,u_{1,V}\otimes D_{1,uV}^h+e_d^2\,d_{1,V}\otimes D_{1,dV}^h\nn
F_{UU}^{\cos 2\phi_h, h-\bar h}&=&
\left[e_u^2\,h_{1,uV}^\perp \otimes H_{1,uV}^{\perp h}+e_d^2\,h_{1,dV}^\perp
\otimes H_{1,dV}^{\perp h}\right]\otimes w_2^\perp+\nn
&&+\frac{2}{Q^2}\left[e_u^2\,u_{1,V}\otimes D_{1,uV}^h+e_d^2\,d_{1,V}\otimes D_{1,dV}^h\right]\otimes w_2\nn
F_{UU}^{\cos \phi_h, h-\bar h}&=&-\frac{2}{Q}
\left\{\left[e_u^2\,h_{1,uV}^\perp \otimes H_{1,uV}^{\perp h}+e_d^2\,h_{1,dV}^\perp
\otimes H_{1,dV}^{\perp h}\right]\otimes w_1^\perp\right.+\nn
&&\left.+\left[e_u^2\,u_{1,V}\otimes D_{1,uV}^h+e_d^2\,d_{1,V}\otimes D_{1,dV}^h\right]\otimes w_1\right\}\label{diffp}
\ee
 In this expressions we've  neglected the terms proportional to $s_{1V}\equiv s_1-\bar s_1$, which are small
 being proportional to $s-\bar s$, on which
as  a strong bound  from neutrino experiments exists, $\vert s-\bar
s \vert \leq 0.025$\,\cite{Soffer}.  We've also neglected the terms proportional to
  $h^\perp_{sV}\equiv h^\perp_{1s}-h_{1\bar s}^\perp$,  which are small due to the positivity condition
   $h^\perp_{sV} \leq s_{1V}$.

In our approach, naturally the TMD densities of the valence quarks $q_V=q-\bar q$, appear. They
 fragment into the the final hadrons and  the TMD  valence-quark FFs appear: $D_{qV}^h=D_{q-\bar q}^h=D_q^{h-\bar h}$
 (not to be confused with favoured FFs!).  We use the  notation:
 \be
 u_{1V}=u_1-\bar u_1,\quad d_{1V}=d_1-\bar d_1,\quad  h_{1,uV}^\perp =h_{1,u}^\perp - h_{1,\bar u}^\perp ,\quad
 h_{1,dV}^\perp =h_{1,d}^\perp - h_{1,\bar d}^\perp
 \ee
 \be
&&D_{1,uV}^h\equiv D_{1,u}^h-D_{1,\bar u}^h,\qquad D_{1,dV}^h\equiv D_{1,d}^h-D_{1,\bar d}^h,\nn
&&H_{1,uV}^{\perp h}\equiv H_{1,u}^{\perp h}-H_{1,\bar u}^{\perp h},
\qquad H_{1,dV}^{\perp h}\equiv H_{1,d}^{\perp h}-H_{1,\bar d}^{\perp h}
 \ee

In stead of the sum over all quark flavours  $q=u,\bar u,d,\bar d, s,\bar s$
  in $d\s^h_p$, in $d\s^{h-\bar h}_p$ we
 have a sum over the two valence $u_V$ and $d_V$ quarks only. The sea quarks do not contribute.

 In addition, the FFs that survive -- $D_{1,uV}^h$, $D_{1,dV}^h$, and
$H_{1,uV}^h$, $H_{1,dV}^h$ couple to the large
 valence quark densities $q_{1,V}$ and $h_{1,qV}^\perp$. The strange-quark TMD-FFs
 $D_{1,sV}^h$ and $H_{1,sV}^h$ are suppressed by the small factor $(s-\bar s)$ and  we safely neglect them.

 \subsection{ on deuterium target}

SU(2) invariance implies that the cross section on a neutron target is obtained from (\ref{diffp}) with the
 replacements of the $u$ and $d$ parton densities:
\be
u_{1V}\leftrightarrow d_{1V},\qquad s_{1V}\to s_{1V},\qquad
h_{1,uV}^\perp\leftrightarrow h_{1,dV}^\perp ,\qquad h_{1,sV}^\perp\to h_{1,sV}^\perp ,\qquad
\ee
 Then,  for the contributions to the  cross section $ d\s_d^{h-\bar h}$ on a deuterium target:
 \be
 d\s_d^{h-\bar h}=d\s_p^{h-\bar h}+d\s_n^{h-\bar h}
 \ee
   we obtain:
  \be
F_{UU}^{h-\bar h}(d=p+n)&=& (u_{1,V}+d_{1,V})\otimes \left(e_u^2\,D_{1,uV}^h+e_d^2\, D_{1,dV}^h\right)\nn
F_{UU}^{\cos 2\phi_h, h-\bar h}(d=p+n)&=& \left(h_{1,uV}^\perp + h_{1,dV}^\perp \right)\otimes
 \left(e_u^2\,H_{1,uV}^{h\perp}+e_d^2\,H_{1,dV}^{h\perp}\right)\otimes w_2^\perp +\nn
&&+\frac{2}{Q^2}\left(\,u_{1,V}+d_{1,V}\right)\otimes \left(e_u^2\,D_{1,uV}^h+e_d^2\, D_{1,dV}^h\right)\otimes w_2\nn
F_{UU}^{\cos \phi_h, h-\bar h}(d=p+n)&=&-\frac{2}{Q}
\left\{\left(h_{1,uV}^\perp + h_{1,dV}^\perp \right)\otimes
 \left(e_u^2\,H_{1,uV}^{h\perp}+e_d^2\,H_{1,dV}^{h\perp}\right)\otimes w_1^\perp +\right.\nn
&&\left.+\left(\,u_{1,V}+d_{1,V}\right)\otimes \left(e_u^2\,D_{1,uV}^h+e_d^2\, D_{1,dV}^h\right)\otimes w_1\right\}
\nn\label{diffd}
\ee

Note that only 2 combinations of TMD valence-quark densities:
$(u_{1,V}+d_{1,V})$ and $\left(h_{1,uV}^\perp + h_{1,dV}^\perp \right)$,
and only 2 combinations of TMD-FFs: $\left(e_u^2\,D_{1,uV}^h+e_d^2\, D_{1,dV}^h\right)$
and $ \left(e_u^2\,H_{1,uV}^{h\perp}+e_d^2\,H_{1,dV}^{h\perp}\right)$ enter.
 In addition, TMD-PDFs and TMD-FFs do not mix and, each one can be parametrized separately.

These expressions are further simplified when the final hadrons are specified.

 \section{The difference cross section with  ${\bf \pi^\pm}$}

 When the final hadrons are $\pi^\pm$, SU(2) invariance  of strong interactions
 implies:
\be
 D_{1,uV}^{\pi^+} \equiv  D_{1,u}^{\pi^+}- D_{1,\bar u}^{\pi^+}=-D_{1,dV}^{\pi^+}\label{SU2}
\ee
and similarly for the Collins FFs $H_{1,q}^{\perp \pi^\pm}$:
\be
H_{1,uV}^{\perp \pi^+} =-H_{1,dV}^{\perp \pi^+}.
\ee
Then from (\ref{diffp}) and (\ref{diffd}) we obtain
 the difference cross sections $d\s^{(\pi^+-\pi^-)}$.
 We present the expressions for a proton and deuterium targets separately.

 \subsection{ on proton target}

From (\ref{diffp}), for the contributions to $d^5\s_p^{\pi^+-\pi^-}$ we obtain:
\be
F_{UU}^{\pi^+ -\pi^-}&=&\left( e_u^2\, u_{1,V}-e_d^2 \,d_{1,V}\right)\otimes D_{1,uV}^{\pi^+}\nn
F_{UU}^{\cos 2\phi_h, \pi^+-\pi^-}&=&
\left(e_u^2\,h_{1,uV}^\perp -e_d^2 \,h_{1,dV}^\perp\right)\otimes H_{1,uV}^{\perp \pi^+} \otimes w_2^\perp+\nn
&&+\frac{2}{Q^2}\,\left( e_u^2 u_{1,V}-e_d^2 d_{1,V}\right)\otimes D_{1,uV}^{\pi^+}\otimes w_2\nn
F_{UU}^{\cos \phi_h, \pi^+-\pi^-}&=&
-\frac{2}{Q}\left\{\left(e_u^2\,h_{1,uV}^\perp -e_d^2 \,h_{1,dV}^\perp\right)\otimes H_{1,uV}^{\perp \pi^+} \otimes w_1^\perp +\right.\nn
&&\hspace*{2cm}\left.+\,\left( e_u^2 \,u_{1,V}-e_d^2\, d_{1,V}\right)\otimes D_{1,uV}^{\pi^+}\otimes w_1\right\}
\ee

 \subsection{ on deuterium target}

From (\ref{diffd}), for the contributions to $d^5\s_d^{\pi^+-\pi^-}$ we obtain:
\be
F_{UU}^{\pi^+ -\pi^-}&=&\left( e_u^2 -e_d^2\right)\left(u_{1,V}+ d_{1,V}\right)\otimes D_{1,uV}^{\pi^+}\nn
F_{UU}^{\cos 2\phi_h, \pi^+-\pi^-}&=&
\left( e_u^2 -e_d^2\right)\left\{\left(h_{1,uV}^\perp + h_{1,dV}^\perp\right)
\otimes H_{1,uV}^{\perp \pi^+} \otimes w_2^\perp+\right.\nn
&&\left.+\frac{2}{Q^2}\,\left(  u_{1,V}+ d_{1,V}\right)\otimes D_{1,uV}^{\pi^+}\otimes w_2\right\}\nn
F_{UU}^{\cos \phi_h, \pi^+-\pi^-}&=&
-\frac{2}{Q}\left( e_u^2 -e_d^2\right)\left\{\left(h_{1,uV}^\perp + h_{1,dV}^\perp\right)
\otimes H_{1,uV}^{\perp \pi^+} \otimes w_1^\perp +\right.\nn
&&\hspace*{2cm}\left.+\,\left(  u_{1,V}+ d_{1,V}\right)\otimes D_{1,uV}^{\pi^+}\otimes w_1\right\}
\ee

It is just one TMD-FF for unpolarized $ D_{1,uV}^{\pi^+}$ and one for polarized quarks
$H_{1,uV}^{\perp \pi^+}$  that enter, which would allow to determine them
 independently of the behaviour of the other TMD's.

\section{The difference cross section with  ${\bf K^\pm}$}

If we consider only charged kaons, we cannot use SU(2)-invariabce as it relates neutron to charged kaons.
However, in order to simplify analysis, the assumption made
in  all analysis of kaon production is that the unfavoured FFs of $d$ and $\bar d$ quarks into $K^+$  are the same:
\be
D_{dV}^{K^+}=H_{dV}^{\perp K^+}=0\label{K}
\ee
Bellow we present  the functions $F_{UU}^{K^+ -K^-}$ in $d\s^{K^+-K^-}$ for
 proton and deuterium targets, using this assumption.

 \subsection{ on proton target}

  From (\ref{diffp}) for the terms $F_{UU}^{K^+-K^-}$ in  $d^5\s_p^{K^+-K^-}$ we obtain:
\be
F_{UU}^{K^+ -K^-}&=& e_u^2 u_{1,V}\otimes D_{1,uV}^{K^+}\nn
F_{UU}^{\cos 2\phi_h, K^+-K^-}&=& e_u^2\left\{h_{1,uV}^\perp \otimes H_{1,uV}^{\perp K^+} \otimes w_2^\perp
+\frac{2}{Q^2}\,u_{1V}\otimes D_{1,uV}^{K^+}\otimes w_2\right\}\nn
F_{UU}^{\cos \phi_h, K^+-K^-}&=&-\frac{2}{Q}\,e_u^2\left\{h_{1,uV}^\perp \otimes H_{1,uV}^{\perp K^+} \otimes w_1^\perp +
\, u_{1,V}\otimes D_{1,uV}^{K^+}\otimes w_1\right\}
\ee

 \subsection{ on deuterium target}

 From (\ref{diffd}) for $d\s_d^{K^+-K^-}$ we obtain:
\be
F_{UU}^{K^+ -K^-}&=& e_u^2 \left( u_{1,V}+d_{1,V}\right)\otimes D_{1,uV}^{K^+}\nn
F_{UU}^{\cos 2\phi_h, K^+-K^-}&=& e_u^2\left\{\left( h_{1,uV}^\perp +h_{1,dV}^\perp \right)
\otimes H_{1,uV}^{\perp K^+} \otimes w_2^\perp
+\right.\nn
&&\hspace*{3.5cm}\left.+\frac{2}{Q^2}\,\left(u_{1,V}+d_{1,V}\right)\otimes D_{1,uV}^{K^+}\otimes w_2\right\}\nn
F_{UU}^{\cos \phi_h, K^+-K^-}&=&-\frac{2}{Q}\,e_u^2\left\{\left(h_{1,uV}^\perp +h_{1,dV}^\perp \right)
\otimes H_{1,uV}^{\perp K^+}
 \otimes w_1^\perp +\right.\nn
&&\hspace*{3.5cm}\left.+\,\left(u_{1,V}+d_{1,V}\right)\otimes D_{1,uV}^{K^+}\otimes w_1\right\}
\ee

Here we summarize some common features of the considered difference cross sections:

{\bf 1.} On a deuterium target, both for $h-\bar h$,  $\pi^+-\pi^-$ and $K^+-K^-$,
always the same combinations of TDM parton densities are measured: $(u_{1,V}+d_{1,V})$ and $(h_{1,uV}^\perp + h_{1,dV}^\perp )$.

{\bf 2.} On a deuterium target, it is always one combination of unpolarized and one of polarized quarks TMD-FFs that enter.
This combination depends on the final hadron -- for charged hadrons it is
$\left(e_u^2\,D_{1,uV}^h+e_d^2\, D_{1,dV}^h\right)$
and $ \left(e_u^2\,H_{1,uV}^{h\perp}+e_d^2\,H_{1,dV}^{h\perp}\right)$, for $\pi^\pm$ it is
$ D_{1,uV}^{\pi^+}$ and $H_{1,uV}^{\perp \pi^+}$,
for $K^\pm$ it is
$ D_{1,uV}^{K^+}$ and $H_{1,uV}^{\perp K^+}$. However, the key point, that it's  always one quantity, remains
which allows to  extract it
  irrespectively from the other TMD-FFs.

{\bf 3.}   Both on proton and deuterium targets, only the  valence quarks  TMD functions  enter
 all difference cross sections.

\section{The  cross section for  ${\bf {\cal K}=K^++K^--2K_s^0}$}

Up to now we considered production of any charged hadrons, $h-\bar h$ and $ \,h=\pi^\pm,\, K^\pm$.
Now we consider production of kaons only.

If in addition to the charged  $K^\pm$ also  neutral kaons $K_s^0=(K^0+\bar K^0)/\sqrt 2$
   are measured,  SU(2) invariance
   of the strong interactions implies that no new FFs are introduced into the cross-sections.  We have:
\be
 &&D_{1u}^{K^+ + K^--2K_s^0}=-D_{1d}^{K^+ +
K^--2K_s^0}={(D_{1u}-D_{1d})}^{K^+ + K^-},\nn
&&D_{1s}^{K^+ + K^--2K_s^0}=
D_{1c}^{K^+ + K^--2K_s^0}=D_{1b}^{K^+ + K^--2K_s^0}=0, \label{SU2kaons}
 \ee
 and similarly for $H_{1q}^{\perp ,h}$.

  We show that, in the difference of charged and neutral kaons production  in SIDIS, $d\s^{\cal K}$:
   \be
d\sigma^{\cal K}=d\sigma ^{K^++K^--2K_s^0}\equiv d\sigma ^{K^+}+d\sigma ^{K^-}-2d\sigma
^{K_s^0}
 \ee
 only one combination of unpolarized TMD-FFs: ${(D_{1u}-D_{1d})}^{K^+ + K^-}$
 and one combination of Collins-functions ${(H_{1u}-H_{1d})}^{\perp ,K^+ + K^-}$
 enter, both, for  proton and deuterium targets. This result
  is obtained under the only assumption of SU(2)-invariance.
 We give the expressions for $d\s^{\cal K}$ on proton and deuterium targets.

\subsection{ on proton target}

Using (\ref{SU2kaons}) for $d\s_p^{\cal K}$ we obtain:
 \be
 F_{UU}^{\cal
K}&=& \left[e_u^2 (u_1+\bar u_1) -e_d^2(d_1+\bar d_1)\right]\otimes
  D_{1,u-d}^{K^++K^-}\nn
F_{UU}^{\cos 2\phi_h, {\cal K}}&=& \left[ e_u^2(h_{1,u}^\perp +h_{1,\bar u}^\perp ) -e_d^2 (h_d^\perp +h_{\bar d}^\perp ) \right]
 \otimes H_{1,u-d}^{\perp , K^++K^-} \otimes w_2^\perp +\nn
 &&+\frac{2}{Q^2}\,\left[e_u^2 (u_1+\bar u_1) -e_d^2(d_1+\bar d_1)\right]\otimes
  D_{1,u-d}^{K^++K^-} \otimes w_2\nn
 F_{UU}^{\cos \phi_h, {\cal K}}&=&-\frac{2}{Q}\,\left\{ \left[ e_u^2(h_{1,u}^\perp +h_{1,\bar u}^\perp )
 -e_d^2 (h_{1,d}^\perp +h_{1,\bar d}^\perp ) \right]\otimes  H_{1,u-d}^{\perp , K^++K^-} \otimes w_1^\perp +\right.\nn
&&\left.\hspace{1cm} +\left[e_u^2 (u_1+\bar u_1) -e_d^2(d_1+\bar d_1)\right]\otimes
  D_{1,u-d}^{K^++K^-} \otimes w_1\right\}
  \ee
  Here we have used the brief notation:
  \be
  D_{1,u-d}^{K^++K^-}&=&{(D_{1u}-D_{1d})}^{K^+ + K^-}\nn
H_{1,u-d}^{\perp , K^++K^-}&=&{(H_{1u}-H_{1d})}^{\perp ,K^+ + K^-}
\ee
\subsection{ on deuterium  target}

Using (\ref{SU2kaons}) for $d\s_d^{\cal K}$ we obtain:
\be
F_{UU}^{\cal K}&=&(e_u^2-e_d^2)  \left(u_1+\bar u_1+d_1+\bar d_1\right)\otimes
  D_{1,u-d}^{K^++K^-}\nn
F_{UU}^{\cos 2\phi_h, {\cal K}}&=& (e_u^2-e_d^2)  \left\{(h_{1,u}^\perp +h_{1,\bar u}^\perp + h_{1,d}^\perp +h_{1,\bar d}^\perp )
 \otimes H_{1,u-d}^{\perp , K^++K^-} \otimes w_2^\perp +\right.\nn
 &&\left.+\frac{2}{Q^2}\,(u_1+\bar u_1+d_1+\bar d_1)\otimes
  D_{1,u-d}^{K^++K^-} \otimes w_2\right\}\nn
 F_{UU}^{\cos \phi_h, {\cal K}}&=&-\frac{2}{Q}\,(e_u^2-e_d^2)
 \left\{(h_{1,u}^\perp +h_{1,\bar u}^\perp +h_{1,d}^\perp +h_{1,\bar d}^\perp )
 \otimes H_{1,u-d}^{\perp , K^++K^-} \otimes w_1^\perp +\right.\nn
 &&\left.+\,(u_1+\bar u_1+d_1+\bar d_1)\otimes
  D_{1,u-d}^{K^++K^-} \otimes w_1\right\}
  \ee

Common for all differences is that  TMD parton densities  factorize from FFs.

\section{Parametrizations and Comments}

Up to now all considerations were general, based only on C and SU(2)-invariance of strong interactions,
 with no assumptions on the parametrizations of the TMD-PDFs and the TMD-FFs. Here we shall summarize the conventionally used
parametrisations  and suggest how they  modify when applied to the considered approach.

There are 4 types of TMDs  for each quark flavour $q=u,\bar u,d , \bar d, s,\bar s$,
 that enter the differential cross sections $d\s^h_N$
of unpolarized SIDIS -- the unpolarized quark densities $q_{1}$ that couple unpolarized
FFs $D_{1,q}^h$, and the
  transversely polarized quarks densities $h_{1,q}^\perp$ that couple to
the  transversely polarized  FFs $H_{1,q}^\perp$. This makes, in total, 24 independent
 quantities for each type of hadrons, that have to be determined.

In the considered here difference cross sections, the 4 types of TDMs are only for the 2 valence-quarks
 $q_V=u_V,\,d_V$ -- the unpolarized valence-quark densities $q_{1V}$ that couple unpolarized valence-quark
FFs $D_{1,q_V}^h$, and the
  transversely polarized valence-quarks densities $h_{1qV}^\perp$ that couple to
the  transversely polarized valence-quarks FFs $H_{1,qV}^\perp$. The independent unknown quantities is reduced in total to at most  8.

Many simplifying assumptions are made in the performed conventional
analysis of $d\s^h_N$: the $x$ and $k_\perp$ ($z$ and $p_\perp$) dependence is  factorized
 with a Gaussian dependence on
 the transverse momenta, no flavour, no $Q^2$,  no $x$ and no $z$-dependencies
 in  the transverse-dependent parts,
 the $Q^2$ evolution is only in the
collinear PDFs and FFs according to the DGLAP equations.
 Here we present the standard parametrizations for the TMD quark densities
 and FFs for all quark flavours,  and discuss how they can be modified for the TMD valence quarks.
 We comment on the advantages of the considered approach.
  We consider the  unpolarized and the transversely polarized quark TMD functions separately.

\subsection{ The $\phi_h$ - independent terms}

{\bf 1.} The TMD parton densities and fragmentation functions with  unpolarized quarks
 are \cite{Anselmino_2005}, $q=u,\bar u, d,\bar d, s,\bar s$:
\be
q_{1}(x,k_\perp , Q^2)&=&q(x,Q^2)\,\frac{e^{-k_\perp^2/<k_\perp^2>}}{\pi<k_\perp^2>} \nn
D_{1,q}^h(z,p_\perp ,Q^2) &=&D_{q}^h(z,Q^2) \,\frac{e^{-p_\perp^2/<p_\perp^2>}}{\pi<p_\perp^2>},
\ee
where $q(x)$ and $D_d^h$ are the collinear PDFs and FFs. The fitting parameters
$<k_\perp^2>$ and $<p_\perp^2>$ are  assumed flavour independent.
 This leads to a Gaussian-type dependence on $P_T^2$, with a $z_h$-dependent width
$\langle P^2_T\rangle$:
\be
\langle P^2_T\rangle = \langle p_\perp^2\rangle  + z_h \langle k_\perp^2\rangle .
\ee

In a very recent analysis \cite{Anselmino_2013}, from a separate fit to multiplicities in the unpolarized SIDIS
data of COMPASS (with charged unidentified hadron $h^\pm$ on a deutron) and HERMES
 (with $\pi^\pm$ and $K^\pm$ on proton and deuterium targets),
 $<k_\perp^2>$ and $<p_\perp^2>$ were determined with good precision.

{\bf 2.}   In the discussed differences only  the 2 valence-quark TMD parton densities $u_{1,V}$ and $d_{1,V}$,
 and the 2 valence-quark TDM-FFs
 $D_{1,uV}^h$ and $D_{1,dV}^h$ enter. They can be parametrized analogously, $q_V=u_V, d_V$:
\be
q_{1,V}(x,k_\perp , Q^2)&=&q_V(x,Q^2)\,\frac{e^{-k_\perp^2/<k_\perp^2>_{qV}}}{\pi<k_\perp^2>_{qV}},\label{fq}\\
D_{1,q_V}^h(z,p_\perp ,Q^2) &=&D_{q_V}^h(z,Q^2) \,\frac{e^{-p_\perp^2/<p_\perp^2>_{qV}}}{\pi<p_\perp^2>_{qV}},\label{Dq}
\ee
where $q_V$ and $D_{qV}^h$ are the  collinear valence PDFs and FFs.
 As multiplicities on proton and deuterium targets provide
2 independent measurements, one could relax the assumption of flavour independence and fit data with  flavour dependent
parameters $<k_\perp^2>_{qV}$ and $<p_\perp^2>_{qV}$, $q_V=u_V,d_V$. This implies that
  the $P_T^2$-dependence will not longer be a simple Gaussian distribution.
 Recently, first studies
 on  flavour dependence of the partonic  transverse momentum  in  unpolarized TMD functions was done
 and interesting results were obtained \cite{flavour}.
 We hope this will help such investigations.

 {\bf 3.}  Measurements  on deuterium target with $h-\bar h$, $\pi^+-\pi^-$ and $K^+-K^-$ final hadrons
 provide 3 independent measurements for the sum of the valence-quark  TMD: $u_{1,V} + u_{1,V}$.

 {\bf 4.}  Measurements  on deuterium target
 always measures only one combination of the unpolarized valence-quark TMD-FFs, which allows to determine it
 without additional assumptions, independently from the other TMD-FFs. The combination  depends on the final hadron
 $h-\bar h$, $\pi^+-\pi^-$ or $K^+-K^-$.

\subsection{ The $\phi_h$ - dependent terms}

{\bf 1.} Using the ansatz of refs.~\cite{Anselmino_09} - \cite{Barone_2010},
  the Boer-Mulders and Collins functions, $h_{1,q}^\perp$ and $H_{1,q}^{\perp, h}$, are most generally
   proportional to the unpolarized TMD PDFs and FFs, respectively:
 \be
   h^\perp_{1,q}(x,k_\perp , Q^2)&=&\rho_q(x)\,\eta (k_\perp ) f_{1,q}(x,k_\perp, Q^2)\nn
 H^{\perp ,h}_{1,q}(z,p_\perp , Q^2)&=&\rho_q^{C}(z)\,\eta^C(p_\perp )\, D_{1,q}^h(z,p_\perp,Q^2)
 \ee
 where $\rho_q(x)$ and $\rho_q^C(z)$, $\eta (k_\perp ) $ and $\eta^C (p_\perp ) $ are new fitting functions.
 Usually the transverse dependent functions $\eta$ and $\eta^C$ are  assumed flavour independent.

  {\bf 2.} Only  2 valence-quark Boer-Mulders densities $h_{1,uV}$ and $h_{1,dV}$,
 and  2 valence-quark Collins functions $H_{1,uV}^h$ and $H_{1,dV}^h$ enter the difference cross sections.
  They can be parametrized analogously:
 \be
  h^\perp_{1qV}(x,k_\perp , Q^2)&=& \rho_{qV}(x) \,\eta_{qV} (k_\perp )q_{1,V} (x,k_\perp,Q^2)\nn
 &=&  \rho_{qV}(x)\,\eta_{qV}(k_\perp)\,q_V (x,Q^2)\,
  \frac{e^{-k_\perp^2/<k_\perp^2>}}{\pi<k_\perp^2>}
 \ee
 and
\be
   H^{\perp ,h}_{1qV}(z,p_\perp , Q^2)&=&\rho_{qV}^{C}(z)\,\eta_{qV}^C(p_\perp )\, D_{1,qV}^h(z,p_\perp,Q^2)\nn
   &=&\rho_{qV}^C(z)\,\eta_{qV}^C(p_\perp )\, D_{qV}^h(z,Q^2) \frac{e^{-p_\perp^2/<p_\perp^2>}}{\pi\,<p_\perp^2>}.
   \ee
where, given the simplicity of the approach, the TMD functions
 $\eta_{qV}(k_\perp)$ and $\eta_{qV}^C (p_\perp)$ can be considered flavour dependent.
   Measurements of the  $\cos 2\phi_h$ (and
  $\cos \phi_h$) asymmetry on proton and deuterium targets  provide 2 independent measurements that would
  allow, in principle,  to determine them separately.

{\bf 3.}  Measurements  on deuterium target with $h-\bar h$, $\pi^+-\pi^-$ and $K^+-K^-$
 provide 3 independent measurements for the
 sum of the valence-quark Boer-Mulders functions: $h^\perp_{1,uV} + h^\perp_{1,dV}$.

 {\bf 4.} Measurements  on deuterium target
 always measures only one combination of valence-quark Collins functions.
 This allows to determine it
  independently from the other TMD-FFs.
 The combination  depends on the final hadron
 $h-\bar h$, $\pi^+-\pi^-$ or $K^+-K^-$.
\vspace{.5cm}

{\bf 5.} Following the same path of arguments, the parametrizations for $D_{1,u-d}^{K^++K^-}$ and $H_{1,u-d}^{K^++K^-}$
are:
\be
D_{1,u-d}^{K^++K^-}(z,p_\perp,Q^2)&=&D_{u-d}^{K^++K^-}(z,Q^2) \,\frac{e^{-p_\perp^2/<p_\perp^2>_{u-d}}}{\pi<p_\perp^2>_{u-d}}\nn
H_{1,u-d}^{K^++K^-}(z,p_\perp,Q^2)&=&\rho_{u-d}^{C}(z)\,\eta_{u-d}^C(p_\perp )\, D_{1,u-d}^{K^++K^-}(z,p_\perp,Q^2)
\ee
Measurements on proton and deuterium targets could determine $<p_\perp^2>_{u-d}$ and $\eta_{u-d}^C (p_\perp)$ independently, without relations between them and with 
  other  TMD fragmentation functions. Note that the collinear FFs $D_{u-d}^{K^++K^-}(z)$  that enter are known solely from the inclusive
 $e^+e^-$ annihilation process: $e^+e^- \to K^\pm +X$, without the assumptions of favoured and unfavoured FFs,  they evolve in $Q^2$ as non-singlets according to the DGLAP equations.

\subsection{  Common  for  all differences}

{\bf 1.}  All differences rely, as known quantities,
on the collinear valence quark PDFs $u_V$ and $d_V$,
  which are the best known parton densities
  (with  2\%-3\% accuracy at $x\lesssim 0.7$),  and on the collinear valence FFs $D^{h,\pi^+,K^+}_{uV}$.
 Very recently,
  $D^{\pi^+}_{uV}$ were determined  with a very good precision,
   directly in a model independent analysis of the HERMES
  data~\cite{LSS}.

 {\bf 2.}
 The   $Q^2$-dependence of the non-singlets $q_V$ and $D_{qV}^h$, that enter the valence-quark parametrizations,
  is relatively simple. This  would make it easier to investigate
 the $Q^2$-dependence in the transverse-momentum dependent part.
     Recently it was found \cite{Anselmino_2013} that a logarithmic $Q^2$-dependence in
$\langle P^2_T\rangle$ improves description of the data.

\section{Conclusions}

We have presented an alternative approach to extracting the TMD parton densities and FFs that enter
the cross section of unpolarized SIDIS.

Based only on  factorization,
 C - invariance  and SU(2) - invariance of strong interactions,
  without any assumptions about PDFs and FFs, we show that
 the difference cross sections of unpolarized SIDIS
$d\sigma^{h^+}-d\sigma^{h^-}$, $d\sigma^{\pi^+}-d\sigma^{\pi^-}$ and $d\sigma^{K^+}-d\sigma^{K^-}$
 are expressed solely in terms of the valence-quark TMD unpolarized densities $q_{1,V}$ and
Boer-Mulders functions $h_{1,qV}^\perp$, and the valence-quark TMD unpolarized fragmentation
 $D_{1,qV}^{h}$ and Collins $H_{1,qV}^\perp$ functions.
If measurements on proton and deuterium targets are fulfilled, a model independent
information about these quantities can be obtained.
Measurements on  a deuterium target, both for $h-\bar h$,  $\pi^+-\pi^-$ and $K^+-K^-$,
 provide information about the sum of the  valence quark TMD densities
$(u_{1,V}+d_{1,V})$ and $(h_{1,uV}^\perp + h_{1,dV}^\perp )$.

If  in addition to  charged kaons
 $K^\pm$, also the neutral $K^0_s$ can be measured, then SU(2) invariance  implies
 that the combination $d\sigma_N^{K^++K^--K^0_s}$, on both proton and deuterium targets, is
  expressed  in terms of only one combination of the TMD FFs
  $(D_{1,u}-D_{1,d})^{K^++K^-}$ and one combination of  Collins functions $(H_{1,u}^\perp-H_{1,d}^\perp)^{K^++K^-}$.

   The suggested measurements of the difference cross sections  provide
    information  only about the TMD valence-quark densities and  FFs, but they
   allow to determine them separately, without imposing  any relations among them or  with other TMD's.
   They present sort of sum rules, 
based on C and SU(2) invariance,  which 
 reduce the contribution of all individual TMD functions in the cross section,   to a contribution only of the valence-quark TMD functions in the difference cross sections.

\section*{Acknowledgements}
I'm very  thankful to Elliot Leader for  careful reading of the manuscript and his useful comments.
The paper is supported by Grant   "HiggsTools" of  Initial Training
Network of 7-th framework  programme of the EC, and  a priority Grant between Bulgaria and JINR-Dubna.

 \end{document}